# BLACK HOLES IN GALACTIC NUCLEI[1]

*The dynamical evidence*

ROELAND P. VAN DER MAREL[2]
*Institute for Advanced Study*
*Olden Lane, Princeton, NJ 08540, USA*

**Abstract.** The dynamical evidence for black holes (BHs) in galactic nuclei is reviewed, with emphasis on recent improvements in spatial resolution, methods for analyzing galaxy spectra and dynamical modeling. M31, M32 and M87 are discussed in some detail.

## 1. Introduction

Most models of the energy production in quasars and active galactic nuclei (AGN) invoke the presence of massive black holes, with $M_{\rm BH} \approx 10^6 - 10^9$ $M_\odot$ (e.g., Rees 1984). The observed number of quasars at high redshifts implies that many of the normal galaxies today must have gone through an active phase in the past. BHs should thus be common in the nuclei of quiescent galaxies as well (Lynden-Bell 1969; Chokshi & Turner 1992).

The gravitational attraction of a BH in a galactic nucleus influences the distribution and dynamics of the surrounding stars and gas. There will be a central power-law cusp in the stellar surface brightness profile (Bahcall & Wolf 1976; Young 1980; Cipollina & Bertin 1994; Quinlan et al. 1994). The stellar and gas velocities (mean or RMS) will be Keplerian, i.e., $v \propto r^{-1/2}$.

Spatial resolution is the main difficulty in detecting these effects. The influence of a nuclear BH extends approximately to a radius $r = GM_{\rm BH}/\sigma^2$, where $\sigma$ is the galaxy's virial velocity. In practice this radius is often $\leq 1''$, of the same order as the seeing FWHM for ground-based observations.

## 2. Overview

Hubble Space Telescope (HST) photometry has shown that most elliptical galaxies have central surface brightness cusps (e.g., Crane et al. 1993; Fer-

---

rarese et al. 1994; Kormendy et al. 1994), consistent with the predictions of models with BHs. However, other physical processes can also lead to high central mass densities in galaxies (e.g., Kormendy 1993). Kinematical data are thus required to firmly establish the presence of BHs.

The gas kinematics in the narrow- and broad-line regions of AGN has been much studied (e.g., Whittle 1994). Gas motions > 1000 km/s can be observed. However, it remains ill-established whether the motions are gravitational, due to in- or outflow, or chaotic (e.g., Osterbrock 1991), mainly because the emitting regions are not spatially resolved from the ground. The motions can thus not be used directly to trace the gravitational potential. This will improve with high spatial resolution spectroscopy from the refurbished HST, like that obtained already for M87 (Section 3).

Ground-based stellar kinematical studies have produced tentative evidence for the presence of BHs in M31, M32, M87, NGC 3115 and NGC 4594 (e.g., Kormendy 1993), but models without a BH have not been convincingly ruled out. The unknown stellar velocity dispersion anisotropy introduces the main uncertainty. Models with an excess of stars on radial orbits predict a central peak in the RMS stellar velocity, very similar to what is predicted in isotropic models with a BH (Binney & Mamon 1982).

The dynamical modeling of kinematical data has recently improved significantly. Traditionally, spherical or isotropic models were used (in fact, our understanding of these models is still improving: Tremaine et al. 1994; van der Marel 1994d). Now, more sophisticated flattened models are available. Those with a distribution function $f(E, L_z)$ have been particularly well studied (e.g., Hunter & Qian 1993; Dehnen & Gerhard 1994; van der Marel et al. 1994b; Qian et al. 1994), but more general models can also be constructed (Dehnen & Gerhard 1993).

Methods for analyzing absorption-line spectra have traditionally assumed the stellar line-of-sight velocity profiles (VPs) to be Gaussian. Recently, new techniques have been developed that allow the actual VP shapes to be measured (e.g., Rix & White 1992; van der Marel & Franx 1993). Van der Marel et al. 1994a,b,c observed the nearby galaxies with suspected BHs, and measured and modeled their VPs. These data contain new information on the dynamical structure of these galaxies, and allow more stringent constraints to be placed on the presence of BHs.

Stellar kinematical HST data, soon to be expected, should provide much new information. If a BH is present in a galaxy, the central VP will have broader wings than a Gaussian. Gaussian VP fits will strongly underestimate the true velocity dispersion (van der Marel 1994d). Modeling and analysis of VP shapes is thus essential. Flattened models will generally be required, since many galaxies show nuclear stellar disks when viewed with HST (van den Bosch et al. 1994).

## 3. Some individual cases

For M31, M32 and M87 much progress has been made recently:

*M31:* Kormendy (1988a) and Dressler & Richstone (1988) interpreted kinematical data for M31 by invoking the presence of a $10^7 - 10^8 M_\odot$ BH. HST photometry by Lauer et al. (1993) showed that M31 has a double nucleus. Bacon et al. (1994) mapped the stellar kinematics of M31 with the two-dimensional TIGER spectrograph at the CFHT. The faintest of the two nuclei is at the center of the bulge isophotes, and is the kinematical center of the (remarkably regular) velocity field. The velocity dispersion peak is offset from both nuclei. Several explanations for the observed nuclear asymmetries have been discussed, but no consensus has yet been reached. The central regions of M31 might not be in dynamical equilibrium, and the case for a BH (or even two BHs) thus remains ambiguous.

*M32:* Models and observations by Tonry (1987) and Dressler & Richstone (1988) suggested the presence of a BH in M32. Lauer et al. (1992b) presented HST photometry of M32 showing an $I \propto r^{-0.5}$ surface brightness cusp, and argued for a BH on the basis of collision- and relaxation-time arguments. Van der Marel et al. (1994b) and Qian et al. (1994) constructed flattened dynamical models with $f = f(E, L_z)$ to interpret the kinematical data along five different slit position angles presented by van der Marel et al. (1994a). These models require the presence of a $1.8 \times 10^6 M_\odot$ BH, and fit all the data, including the VP deviations from Gaussians, with remarkable accuracy. This does not rule out alternative models, but does put the BH case on much stronger footing. HST observations should provide important new information. The BH model predicts a central velocity dispersion of 129 km/s with the 0.1″ square aperture of the HST/FOS. The central velocity dispersion measured from the ground is only 85 km/s.

*M87:* HST photometry by Lauer et al. (1992a) showed that M87 has an $I \propto r^{-0.26}$ surface brightness cusp. This confirmed photometry by Young et al. (1978), who invoked a $2.6 \times 10^9 M_\odot$ BH as explanation. Sargent et al. (1978) invoked a BH to fit the stellar kinematics. In the 1980's this evidence was demonstrated to be ambiguous. Van der Marel (1994c) showed that the best ground-based stellar kinematical data can still be fit equally well with models with a $3 \times 10^9 M_\odot$ BH, as with models without a BH. He also detected very rapid ionized gas motions near the nucleus of M87, suggested the motions to be gravitational, and demonstrated that the gas motions then imply a $3 \times 10^9 M_\odot$ BH. Beautiful HST observations of the ionized gas have confirmed this hypothesis: the gas lies in a disk (Ford et al. 1994), and rotates with an amplitude of 500 km/s at 0.3″ (Harms et al. 1994). This, almost unambiguously, implies the presence of a $2.4 \times 10^9 M_\odot$ BH. Since M87 is a known active galaxy with an optical synchrotron jet, it does not answer the question whether quiescent galaxies also have BHs.

Support for this work was provided by NASA through a Hubble Fellowship, #HF-1065.01-94A, awarded by the Space Telescope Science Institute which is operated by AURA, Inc., for NASA under contract NAS5-26555.

## References

Bacon R., Emsellem E., Monnet G., Nieto J.-L., 1994, A&A, 281, 691
Bahcall J.N., Wolf R.A., 1976, ApJ, 209, 214
Binney J.J., Mamon G.A., 1982, MNRAS, 200, 361
Chokshi A., Turner E.L., 1992, MNRAS, 259, 421
Cipollina M., Bertin G., 1994, A&A, 288, 43
Crane P. et al. 1993, AJ, 106, 1371
Dehnen W., Gerhard O.E., 1993, MNRAS, 261, 311
Dehnen W., Gerhard O.E., 1994, MNRAS, 268, 1019
Dressler A., Richstone D.O., 1988, ApJ, 324, 701
Ferrarese L., van den Bosch F.C., Jaffe W., Ford H.C., O'Connell R., 1994, AJ, in press
Ford H.C. et al., 1994, ApJ Letters, submitted
Harms R.J. et al., 1994, ApJ Letters, submitted
Hunter C., Qian E., 1993, MNRAS, 262, 401
Kormendy J., 1988a, ApJ, 325, 128
Kormendy J., 1993, in 'The Nearest Active Galaxies', eds. Beckman J.E., Netzer H., Colina L., Madrid
Kormendy J. et al., 1994, to appear in 'Dwarf Galaxies', ed. Meylan G., ESO, Garching
Lauer T.R. et al., 1992a, AJ, 103, 703
Lauer T.R. et al., 1992b, AJ, 104, 552
Lauer T.R. et al., 1993, AJ, 106, 1436
Lynden-Bell D., 1969, Nature, 223, 690
Osterbrock D.E., 1991, Rep. Prog. Phys., 54, 579
Qian E., de Zeeuw P.T., van der Marel R.P., Hunter C., 1994, MNRAS, submitted
Quinlan G.D., Hernquist L., Sigurdsson S., 1994, ApJ, submitted
Rees M.J., 1984, ARA&A, 22, 471
Rix H.W., White S.D.M., 1992, MNRAS, 254, 389
Sargent W.L.W., Young P.J., Boksenberg A., Shortridge K., Lynds C.R., Hartwick F.D.A., 1978, ApJ, 221, 731
Tonry J.L., 1987, ApJ, 322, 632
Tremaine S. et al., 1994, AJ, 107, 634
van den Bosch F.C., Ferrarese L., Jaffe W., Ford H.C., O'Connell R., 1994, AJ, in press
van der Marel R.P., Franx M., 1993, ApJ, 407, 525
van der Marel R.P., Rix. H-W., Carter D., Franx M., White S.D.M., de Zeeuw P.T., 1994a, MNRAS, 268, 521
van der Marel R.P., Evans N.W., Rix H-W., White S.D.M., de Zeeuw P.T., 1994b, MNRAS, in press
van der Marel R.P., 1994c, MNRAS, in press
van der Marel R.P., 1994d, ApJ, 432, L91
Whittle M., 1994, in 'Mass-Transfer Induced Activity in Galaxies', ed. Shlosman I., Cambridge University Press, Cambridge, p.63
Young P., Westphal J.A., Kristian J., Wilson C.P., Landauer F.P., 1978, ApJ, 221, 721
Young P.J., 1980, ApJ, 242, 1232